\documentclass{article}
\usepackage{spconf,amsmath,graphicx}
\usepackage{pifont}
\usepackage{stfloats}
\usepackage{multirow}
\usepackage{booktabs}
\usepackage{diagbox}
\usepackage{amssymb}
\usepackage{url}
\usepackage[ruled,linesnumbered]{algorithm2e}
\newcommand{\pluseq}{\mathrel{+}=}

\ninept 

\newcommand\blfootnote[1]{%
  \begingroup
  \renewcommand\thefootnote{}\footnote{#1}%
  \addtocounter{footnote}{-1}%
  \endgroup
}

\title{BER: Balanced Error Rate For Speaker Diarization}
\name{Tao Liu, Kai Yu $^*$}
\address{MoE Key Lab of Artificial Intelligence, AI Institute, X-LANCE Lab, Shanghai Jiao Tong University}
\begin{document}
%
\maketitle
\begin{abstract}
DER is the primary metric to evaluate diarization performance while facing a dilemma: the errors in short utterances or segments tend to be overwhelmed by longer ones. Short segments, e.g., `yes' or `no,' still have semantic information. Besides, DER overlooks errors in less-talked speakers. Although JER balances speaker errors, it still suffers from the same dilemma. Considering all those aspects, duration error, segment error, and speaker-weighted error constituting a complete diarization evaluation, we propose a Balanced Error Rate ({\em BER}) to evaluate speaker diarization. First, we propose a segment-level error rate ({\em SER}) via connected sub-graphs and adaptive IoU threshold to get accurate segment matching. Second, to evaluate diarization in a unified way, we adopt a speaker-specific harmonic mean between duration and segment, followed by a speaker-weighted average. Third, we analyze our metric via the modularized system, EEND, and the multi-modal method on real datasets. {\em SER} and {\em BER} are publicly available at \text{https://github.com/X-LANCE/BER}.
\end{abstract}

\blfootnote{$^*$ Kai Yu is the
corresponding author.}

%
\section{Introduction}
\label{sec:intro}

Speaker diarization~\cite{park2021review,aronowitz2020new} identifies the talkers and their talking duration, solving the problem of `who spoke when.' Speaker diarization is often used as a pre-processing step in audio tasks, and it has several application scenarios: meeting, telephone recording, etc. With the development of speaker diarization, datasets and methods in speaker diarization have shown new trends. For diarization datasets, datasets \cite{ryant2019second,chung2020spot,liu22t_interspeech} become more in line with real scenarios, which consist of spontaneous speeches, diverse sources, and so on. Those features make a large variance in speaker number and speech duration, especially shorter utterances, which is shown in Table \ref{tab:dataset_metrics}. For diarization methods, joint optimized methods,like VBx-based methods\cite{valente2005variational, kenny2008bayesian, landini2022bayesian} or End-to-end neural diarization (EEND)\cite{fujita2019end, kinoshita2021advances, yang22r_interspeech}, and multi-modal methods emerge\cite{he22c_interspeech,liu22t_interspeech}. Those approaches can handle overlapped speech and short utterances well.

However, conventional evaluation metrics\cite{nist2009, ryant2019second} can not evaluate those features well. Diarization error rate (DER)\cite{nist2009}, and Jaccard error rate (JER)\cite{ryant2019second} have a common issue: they evaluate from the duration view, which will cause a phenomenon that errors by short utterances, e.g., less than one second, are often overlooked because the longer ones occupy most to the duration error. Compared with DER, JER can evaluate speaker-weighted duration error. Conversational diarization error rate (CDER)\cite{cheng2022conversational} proposes a metric to alleviate the issue by evaluating from the segment view. The segment means the segmentation from the reference or the hypothesis and may be different from the utterance. In some circumstances, they are the same. CDER counts all unmatched segment numbers via intersection over union (IoU) matching. But, to allow arbitrary segmentation of the hypothesis, CDER merges the adjacent segments with the same speakers. This merging operation will lead to an unexpected bias if the speech interval is large. Besides, a fixed IoU threshold strategy in CDER will lead to a higher tolerance on longer segment lengths. Segment-level metric remains to be improved.






\begin{table}[t]

  \caption[Results]{ We propose Balanced Error Rate (\textbf{{\em BER}}) to evaluate speaker diarization in speaker-weighted, duration, and segment error. }
  \centering
  \resizebox{0.8\linewidth}{!}{
  \begin{tabular}{@{}l| cccc@{}} \toprule
      Method       &  Speaker-weighted &  Duration Error  & Segment Error  \\ \hline

DER\cite{nist2009}  &   &  \ding{51}     \\ 
JER \cite{ryant2019second} &  \ding{51} & \ding{51} &     \\ 
CDER\cite{cheng2022conversational}&   &  & \ding{51}   \\ 
{\em SER}&   &  & \ding{51}   \\ 
\midrule
\textbf{{\em BER}}&  \ding{51} & \ding{51} & \ding{51} \\ 
\bottomrule

  \end{tabular}
  }
    \label{tab_compasion}
\end{table}






From the above analysis, we can find that DER, JER, and CDER evaluate speaker diarization from limited views, and CDER has improvement rooms. So in this paper, we first proposed an improved segment-level metric: {\em SER}. Based on the metric, we offer {\em BER} a balanced error rate to cover all aspects considered by DER, JER, and SER. The comparison is shown in Table \ref{tab_compasion}. We hope {\em BER} forms a comprehensive metric for the speaker diarization community. 

Our contributions are summarized as follows.

\begin{itemize}
  \item [1.] 
  \textbf{{\em SER}}. Segment-level error rate to evaluate the segment error. To accurately evaluate the segment error, we introduce connected sub-graphs for arbitrary segmentation and IoU adaptation strategy to control offset tolerance.
  
  \item [2.] \textbf{{\em BER}}. Balanced error rate to evaluate speaker-weighted error, duration error, and segment error in a unified way.
  
  \item [3.]\textbf{Extensive analysis and experiments.} We evaluate our metric on several dataset including audio-only and audio-visual datasets. Besides, we conduct the experiment on several methods: modularized system, VBx-based system, and EEND. The results reflect that {\em BER} can provide a complete analysis.
  
\end{itemize}


\section{Related Works}
\label{sec:related_works}


\begin{table}[t]

  \caption[Dataset statistics]{Dataset statistics. \textbf{\#spk.}: The min/average/max speaker number per video. \textbf{spk. speech std. (s)}: Speaker speech standard deviation in seconds. \textbf{ segment duration (s)}:  25\%, 50\% and 75\% percentiles of the segment length in seconds. Datasets vary in those aspects, and conventional metrics can not reflect those aspects well. }
  \label{tab:dataset_metrics}
  \centering
  \resizebox{1.0\linewidth}{!}{
  \begin{tabular}{@{}l |ccc@{}} \toprule
             
             & \#spk. & spk. speech std. (s) & segment duration (s)\\  \hline
AMI\cite{mccowan2005ami} & 3 / 3.99 / 5 & 340 &   0.47 / 1.52 / 4.51 \\ 
CALLHOME\cite{callhome}  &2 / 2.57 / 7 & 49  & 0.73 / 1.53 / 2.94 \\ 
DIHARD2\cite{ryant2019second}  & 2 / 6.78 / 14 & 111 & 0.64 / 1.29 / 2.34 \\ 
VoxConverse\cite{chung2020spot} & 1 / 5.55 / 21  & 135 & 1.12 / 3.16 / 8.73\\ 
MSDWild\cite{liu22t_interspeech}  & 2 / 2.73 / 10 & 80 & 0.75 / 1.47 / 3.13 \\

    \bottomrule
  \end{tabular}
}
\end{table}

    


Diarization error rate (DER) proposed by NIST is the standard scoring metric in speaker diarization datasets\cite{mccowan2005ami, callhomeclassic, liu22t_interspeech} and challenges\cite{ryant2019second, ryant2020third}. DER is the summary of false alarm, missed speech, and speaker confusion time to the reference time. DER is widely used in speaker diarization because this metric is straightforward and intuitive. However, DER is less sensitive to short duration. There are two reasons accounting for this. First, the short-duration occupation is naturally less than the longer ones, resulting in more punishment in errors of longer ones. Second, $collar$ is a time option in DER. If $collar$  is set to more than zero, the period within $collar$ size before and after the segment boundary will be abandoned in evaluation. This option was originally designated to avoid manual labeling noise near the boundaries, but a segment duration with less than two $collar$ will also be excluded from evaluation. The DER formulation is shown in Equation \ref{der} where $\text{FA}_\text{all}$, $\text{MS}_\text{all}$, $\text{SC}_\text{all}$ represents total false alarm, missed and speaker confusion duration, respectively. $\text{REF}_{\text{all}}$ is the total reference duration. The DER result is the percentage of all error duration divided by the reference duration.

\begin{equation}
\text{DER}= \frac{\text{FA}_{\text{all}} + \text{MS}_{\text{all}} + \text{SC}_{\text{all}}}{ \text{REF}_{\text{all}}}
\label{der}
\end{equation}

Jaccard error rate(JER)\cite{ryant2019second}, developed in the second DIHARD challenge, evaluates diarization via speaker-weighted duration error. For each reference speaker $m$, $\text{JER}_m$ is a false alarm and missed speech time to the union of reference and hypothesis speech. For each $\text{JER}_i$, the range is from zero to one, representing the degree from all missed to perfect matching. The final JER is the mean of all $\text{JER}_m$, which means that JER equally treats each speaker's errors, even when the speaking time of one speaker is short. Equation \ref{jer} shows the formulation. The average for each speaker gets the ability to measure speaker discrimination to JER. But for each speaker, JER still suffers the same problem as DER: the errors in longer segments overwhelm the short ones.
\begin{equation}
\text{JER}= \frac{1}{M} \sum_{m=1}^{M} \frac{\text{FA}_{m} + \text{MS}_{m}}{\text{UNION}_{m}}
\label{jer}
\end{equation}

Based on short utterances containing semantic information, Conversational diarization error rate(CDER)\cite{cheng2022conversational} evaluate diarization from segment level, which increases the weight for short-phase segments. The calculation pipeline for CDER contains three main steps: merging adjacent segments with the same speaker, optimal mapping, and IoU matching. After IoU matching, the number of all unmatched segments is regarded as $\text{\#error segs}$. CDER formation is shown in Equation \ref{cder}. However, CDER has two drawbacks. First, the merging operation will cause inaccurate results if there are too many speaking gaps within segments. Second, CDER uses a fixed IoU threshold, which is problematic. The default IoU is 0.5, which allows a one-third offset to the reference duration. However, different segment lengths should have varying IoU sensitivity, e.g., the IoU threshold becomes large for longer utterances and small for short ones.

\begin{equation}
\label{cder}
\text{CDER}= \frac{\text{\#error segs}}{\text{\#REF segs}}
\end{equation}


\section{Metric Description}
\label{sec:metric_description}

This section will illustrate our {\em SER} and {\em BER} metrics in detail.

\textbf{Stage 1. Optimal matching}. Optimal matching is to assign speaker identities of reference to the prediction segments due to permutation problems in speaker label assignment. This forms a bipartite graph matching problem which is often saved by Hungarian algorithm\cite{kuhn1955hungarian}. Optimal matching is a standard stage for diarization metrics like DER\cite{nist2009}, JER\cite{ryant2019second}, and CDER\cite{cheng2022conversational}, and we also use the Hungarian algorithm to assign speaker labels. If the speaker number of the hypothesis is large than the reference, the spare part is not matched. JER ignores this part and does not calculate its error. For {\em SER}, we follow JER and do not calculate this part, but in {\em BER}, this part is denoted as $E^{\text{Speaker}^{\text{FA}}}$, representing errors caused by false alarm speakers. Errors caused by false alarm speakers can not be ignored if a system predict too much candidates, especially in EEND that can not handle arbitrary speaker number well. $E^{\text{Speaker}^{\text{FA}}}$ is the harmonic mean of duration and segment error rate of false alarm speakers.

To ensure one-to-one speaker mapping for the following steps, we add an empty hypothesis ($\varnothing$) to fill the gap if the speaker number of the reference is large than the hypothesis. This step is represented by Algorithm \ref{alg:algorithm1} (line 2 to 13) and the $E^{\text{Speaker}^{\text{FA}}}$ part is formulated in line 33.



\textbf{Stage 2. Speaker-specific duration and segment errors}. 
We calculate duration and segment errors for each one-to-one speaker mapping.

\textbf{Duration errors.} Duration rate of $\text{speaker}_s$ ($E_s^\text{DUR}$) is the sum of false alarm ($\text{FA}^\text{DUR}$) and missed duration ($\text{MS}^\text{DUR}$) to the reference duration ($\text{REF}^\text{DUR}$), shown in Equation \ref{ssdur}. Different from JER, our denominator part is the reference duration not the union duration ($\text{UNION}$) which is also denoted in Algorithm \ref{alg:algorithm1} (line 16).

\begin{equation}
E_s^\text{DUR}= \frac{\text{FA}^\text{DUR} + \text{MS}^\text{DUR}}{\text{REF}^\text{DUR}}
\label{ssdur}
\end{equation}

\begin{figure}
    \centering
    \includegraphics[width=1.0\linewidth]{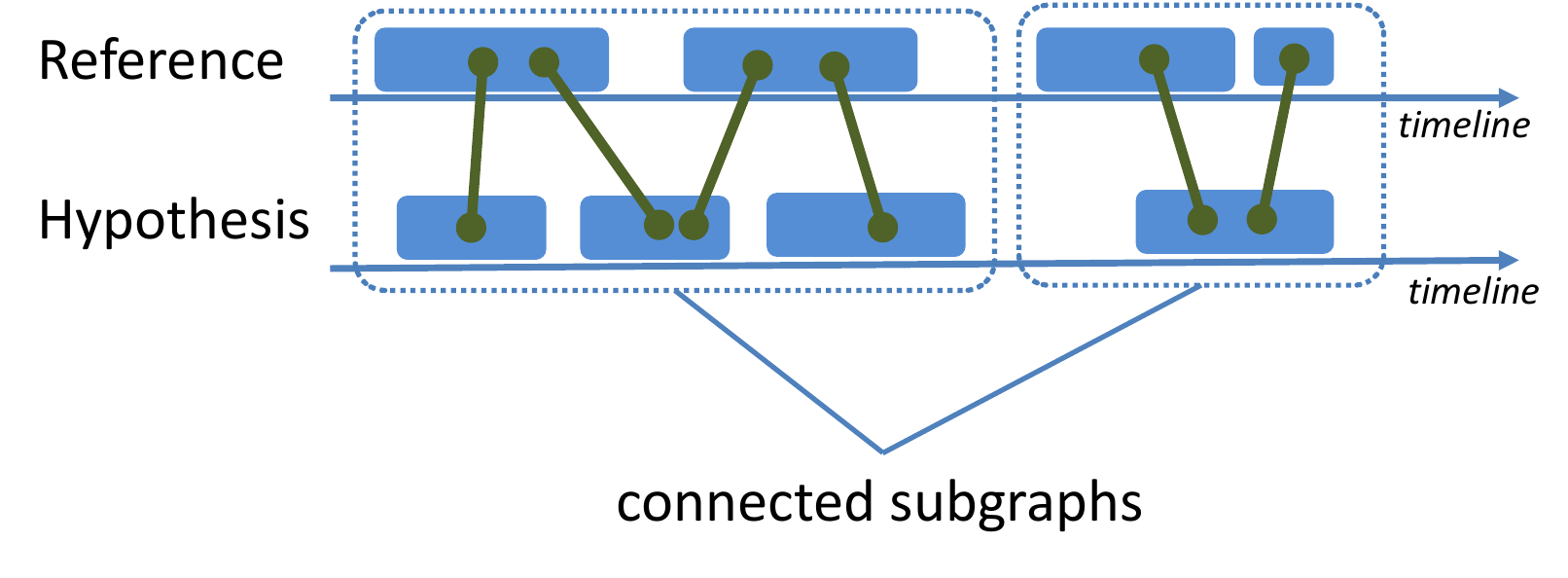}
    \caption{Graph-based segment matching examples. Instead of IoU matching between segments, we run IoU matching under the connected sub-graph. In the graph, nodes are the segments, and edges are the relation between the reference and hypothesis node. The edge exists if there exists an overlap. Nodes and edges in the dashed line mean a connected sub-graph. Via this strategy, our segment-level metric can handle arbitrary segmentation.}
    \label{fig:graph_based_segments}
\end{figure}

\textbf{Segment errors.} 
Unlike direct merging adjacent speakers in CDER\cite{cheng2022conversational}, we adopt a graph-based segment-matching strategy. Specifically, we build a graph to formulate the relations between reference and hypothesis segments. In the graph, the node is the segment or the utterance. If there exists an overlap between the reference and hypothesis, we assign an edge between them. After the graph is constructed, we calculate the connected sub-graphs. We adopt an IoU matching strategy in the reference and hypothesis nodes in each connected sub-graph. It is noted that gaps between nodes are not merged. If IoU is larger than a threshold, we consider that nodes in this sub-graph are connected, which means the reference and hypothesis segments or utterances are matched. Otherwise, the reference segment number in this sub-graph will be considered the error. Through connected sub-graph strategy and only considering reference segment number, segment-level errors can be calculated in arbitrary hypothesis segmentation. This part is consisted with Algorithm \ref{alg:algorithm1} (line 17 - 25) and an illustrated example is shown in Figure \ref{fig:graph_based_segments}. Isolated nodes (without overlapping) in the reference are also considered errors. 

\begin{equation}
E^\text{SEG}_{s} = \frac{\text{\#error segs}}{\text{\#REF segs}}
\label{ssseg}
\end{equation}

For IoU matching strategy, we utilize an adaptive IoU threshold which depends on reference segment duration ($\text{DUR}$) and number ($\text{\#NUM}$). Specifically, we add prior information, $collar$, in this IoU adaption: the offset of the prediction segments must be lower than the size: $2*collar*\text{\#NUM}$. The lower bound($lb$) in this formulation is to prevent too much offset for short segments. The formulation is shown in Equation \ref{iou}. 

\begin{equation}
\resizebox{0.8\linewidth}{!}{ $
 \text{IoU}_\text{Adaption}(\text{DUR},\text{\#NUM}) = \\ 
 max(\frac{\text{DUR}-2*collar*\text{\#NUM}}{\text{DUR}+2*collar*\text{\#NUM}}, lb) $}
 \label{iou}
\end{equation}

\textbf{Stage 3. Speaker-specific harmonic errors}. 

After speaker-specific duration error ($E^\text{DUR}_{s}$) and segment error ($E^\text{SEG}_{s}$) are calculated, we calculate the harmonic mean ($E_s$) between them, which is shown in Equation \ref{sserrors}. Compared with arithmetic mean, harmonic mean prefers to the better result between duration and segment error. In addition, $\text{eps}$ is used here to avoid errors being zero.
\begin{equation}
E_s = \frac{2}{\frac{1}{E^\text{DUR}_{s} + \text{eps}} + \frac{1}{E^\text{SEG}_{s} + \text{eps}}  } - \text{eps} 
\label{sserrors}
\end{equation}

\textbf{Stage 4. Speaker-weighted errors}. 

Then speaker-weighted errors can be calculated by the average of all speaker-specific errors, which is same to JER. This part is shown in Equation \ref{swerror} and Algorithm \ref{alg:algorithm1} (line 35).

\begin{equation}
E^{\text{Speaker}^{\text{REF}}} = \frac{1}{M}\sum_{s=1}^M{E_s}
\label{swerror}
\end{equation}

\textbf{Stage 5. {\em SER} and {\em BER}}. 

Finally, we can get the final {\em SER} and {\em BER}. {\em SER} is the total number of error segments to the reference segments. We do not use hypothesis because its segmentation will affect the result. {\em BER} is the summary of speaker-weighted and false alarm speaker errors. {\em SER} and {\em BER} is shown in Equation \ref{overall_ser} and \ref{overall_ber}, respectively.


\begin{equation}
\label{overall_ser}
\begin{split}
        \text{SER} = \frac{\text{\#error segs}}{\text{\#REF segs}}
\end{split}
\end{equation}

\begin{equation}
\label{overall_ber}
\text{BER} =  E^{\text{Speaker}^{\text{REF}}} +  E^{\text{Speaker}^{\text{FA}}}
\end{equation}



\begin{algorithm}[!hpt]
	\caption{Pseudo code for SER and BER}
	\label{alg:algorithm1}
	\KwIn{$S_i^{REF} = \{U_m\}_{m=1}^M , i \in \mathbb{N}^+  $; $S_j^{HYP} = \{U_n^{'}\}_{n=1}^N , j \in \mathbb{N} $; \tcp{ $S_i^{REF}$ and $S_j^{HYP}$ represents reference speaker $i$'s and hypothesis speaker $j$'s utterances set respectively. $U$ represents utterance or segment.}}
	\KwOut{ $\text{SER}$ and $\text{BER}$.}  
	\BlankLine
	 Init: $\text{E}_{S^\text{REF}}^\text{TOTAL} = list()$, $\text{FA\_error\_duration} = 0$, $\text{global\_REF\_duration} = 0$, $\text{\#FA\_error\_segs} = 0$, $\text{\#global\_error\_segs} = 0$, $\text{\#global\_REF\_segs} = 0$ \; 
	 operate optimal mapping and gets matched sets: $\text{Set}^{\text{M}}=\{(S_i,S_j), \cdots\}$ and unmatched sets: $\text{Set}^\text{FA,REF} =\{S_i, \cdots\}$,  $\text{Set}^\text{FA,HYP} =\{S_j, \cdots\}$;
	\tcp{$(S_i,S_j)$ means a optimal mapping result between $\text{speaker}_i$ and $\text{speaker}_j$}
	
	\If{$\text{Set}^\text{FA,HYP} \notin \varnothing $}{
  \tcp{ FA means false alarm speaker caused by optimal mapping}
	\ForEach{$S_i^{\text{FA}} \in \text{Set}^\text{FA,HYP}$}{
			$\text{FA\_error\_duration} \pluseq$ total duration of $S_i^{\text{FA}} $ \;
			$\text{\#FA\_error\_segs} \pluseq$ total segment number of $S_i^{\text{FA}} $ \;
		}
	}

	\If{$\text{Set}^\text{FA,REF} \notin \varnothing $}{
	\ForEach{$S_i^{\text{FA}} \in \text{Set}^\text{FA,REF}$}{
			add ($S_i^{\text{FA}}, \varnothing$) to $\text{Set}^{\text{M}}$\tcp{ ensure one-to-one mapping for reference speaker } 
			}
			
		} 
	$s = 0$  ;  \tcp{ reference speaker index}
    
	\ForEach{ $(S_i,S_j) \in \text{Set}^{\text{M}}$}{
  \tcp{calculate speaker-specific errors}
	     
	    $E^\text{DUR}_{s} = \frac{\text{FA}^\text{DUR} + \text{MS}^\text{DUR}}{\text{REF}^\text{DUR}}$	\; \tcp{duration errors}
            \text{\#error\_segs} = 0 , \text{\#REF\_segs} = 0 \;
	    calculate connected subgraphs $G$ \;
	    \ForEach{ $\text{REF nodes}$ and $\text{HYP nodes}$ in G  }{
     $\text{\#REF\_segs} \pluseq \text{\#REF nodes} $
     $\text{IoU} = \frac{\text{REF nodes}^\text{DUR} \cap \text{HYP nodes}^\text{DUR}}{\text{REF nodes}^\text{DUR} \cup \text{HYP nodes}^\text{DUR}}$ \;
\If{$ \text{IoU} < \text{IoU}_\text{Adaption}$}{$\text{\#error\_segs} \pluseq \text{\#REF nodes} $}
	    }

      $E^\text{SEG}_{s} = \frac{\text{\#error segs}}{\text{\#REF segs}}$	\; \tcp{segment errors}

    $E_s = harmonic\_mean(E^\text{DUR}_{s}, E^\text{SEG}_{s}) $ \; 
        \tcp{$\text{speaker}_s$'s error }


    $ \text{\#global\_error\_segs} \pluseq \text{\#error\_segs} $ \;

    $ \text{\#global\_REF\_segs} \pluseq \text{\#REF\_segs} $ \;

    $ \text{global\_REF\_duration} \pluseq $ total duration of $S_i$  \;
	   $s \pluseq 1$ \;
	}
$\text{SER} = \frac{\text{\#global\_error\_segs}}{\text{\#global\_REF\_segs}}$ \;

$E^\text{DUR}_{S^\text{FA}} = \frac{\text{FA\_error\_duration}}{\text{global\_REF\_duration}} , E^\text{SEG}_{S^\text{FA}} = \frac{\text{\#FA\_error\_segs}}{\text{\#global\_REF\_segs}} $ \; 
$ E^{\text{Speaker}^{\text{FA}}}= harmonic\_mean(E^\text{DUR}_{S^\text{FA}}, E^\text{SEG}_{S^\text{FA}} ) $ \; 

  $ E^{\text{Speaker}^{\text{REF}}}  = \frac{1}{M} \sum_{s=1}^M{E_s}$ \;
 $\text{BER} = E^{\text{Speaker}^{\text{REF}}} +  E^{\text{Speaker}^{\text{FA}}}$ \;

\Return SER, BER

\end{algorithm}

\section{Experiments}
\subsection{Setups}


All metrics are evaluated with overlapped speech and no forgiveness collar for all duration errors. We use IoU $0.5$ for CDER and adaption IoU with a lower bound ($lb$) IoU $0.5$ for {\em SER} and {\em BER}. It is noted that $collar$ used in segment-level metrics, {\em SER} and {\em BER}, is intended for segment-level IoU threshold, not for boundary collar. For the evaluating dataset, we test our metric on
AMI\cite{mccowan2005ami} (Mix-Headset) test set, CALLHOME\cite{callhome} (LDC2001S97, Disk-8) part \uppercase\expandafter{\romannumeral2}, DIHARD \uppercase\expandafter{\romannumeral2}\cite{ryant2019second} test set, VoxConverse\cite{chung2020spot} test set and MSDWild\cite{liu22t_interspeech} few-talker set. For CALLHOME\cite{callhome}, part \uppercase\expandafter{\romannumeral1} and part \uppercase\expandafter{\romannumeral2} splitting follows Kaldi Callhome diarization recipe. 





\begin{figure}
    \centering
    \includegraphics[width=1.0\linewidth]{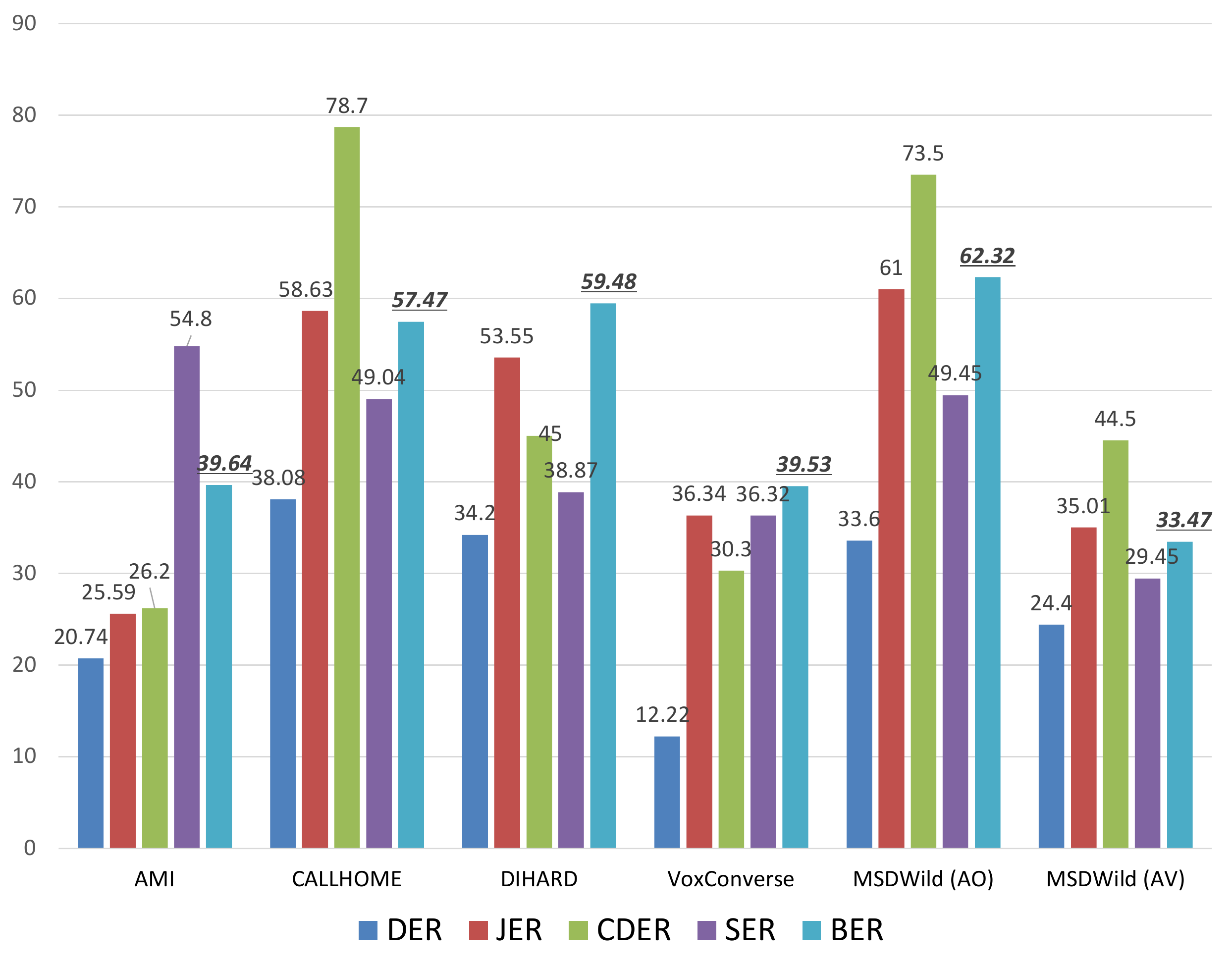}
    \caption{Overview about our proposed {\em SER} and {\em BER} compared with other metrics. Best view in color. AO and AV in MSDWild dataset represents for audio-only and audio-visual respectively. The {\em BER} result is marked with the underline. All results are converted into percentage.}
    \label{fig:overall_comparsion}
\end{figure}

\subsection{Experiments}

\subsubsection{Metric comparison across different datasets }
First, we compare our proposed metrics, {\em SER} and {\em BER}, with other metrics on several publicly available datasets. The comparison result is shown in Figure \ref{fig:overall_comparsion}. All results except for MSDWild (AV) are generated by Pyannote~\cite{Bredin2020}. For Pyannote, we adopt a modularized pipeline: segmentation, ECAPA-TDNN-based\cite{desplanques2020ecapa} embedding extractor, and agglomerative hierarchical clustering (AHC). MSDWild (AV), taking videos and audio as the input, utilizes a multi-modal diarization system, and we follow experiment settings from \cite{liu22t_interspeech}.

Compared with other metrics, {\em BER} considers all aspects: speaker-weighted, duration, and segment error. For example, in AMI, JER is low while {\em BER} is high. We find that there are plenty of false alarm segments. That segment duration is short while the total number is large, leading to a high {\em SER} and {\em BER}. In the MSDWild, benefiting from the visual, the method can predict speech duration more accurately, which reduces DER by 27\% but {\em BER} by 46\%. Those examples indicate that our proposed {\em BER} can provide an overall evaluation for system comparison.


\begin{table}[!hpt]

  \caption[Results]{Metric comparison across different systems on CALLHOME part \uppercase\expandafter{\romannumeral2}.}

  \centering
  \resizebox{1.0\linewidth}{!}{
  \begin{tabular}{@{}l ccccccccc@{}} \toprule
      & &&&& \multicolumn{3}{c}{\textbf{BER}} \\\cmidrule(r){6-8}
       \textbf{Method}       & \textbf{ DER} & \textbf{JER} & \textbf{CDER} & \textbf{SER} & $\text{Speaker}^{\text{REF}}$ & $\text{Speaker}^{\text{FA}}$  &\textbf{Total}  \\ \hline

Modularized system  & 27.68	&48.51	&61.2	&44.95	&49.19	&0.47&	\underline{49.66}  \\ 
VBx\cite{landini2022bayesian} & 21.06	&33.84	&27.7&	36.53&	36.61	&0.42&	\underline{37.02} \\
EEND-VC\cite{kinoshita2021advances} & 23.48	&29.03	&23.3	&23.3&	28.2	&1.15&	\underline{29.35}\\

\midrule

  \end{tabular}
  }
    \label{tab:unbounded_speakernum}
\end{table}

\subsubsection{Metric comparison across different systems}

We also run our metrics on three diarization systems: Modularized system, VBx\cite{landini2022bayesian} and EEND-VC\cite{kinoshita2021advances}, which is shown in Table \ref{tab:unbounded_speakernum}. Our modularized system contains oracle VAD, ECAPA-TDNN-based\cite{desplanques2020ecapa} embedding extractor, and spectral clustering. For a fair comparison, both the modularized system and VBx use oracle VAD, VoxCeleb-based\cite{chung2018voxceleb2} training corpus for speaker embedding, and do not handle overlap speeches. Compared with the modularized system, VBx, a VB-HMM-based diarization algorithm, utilizes HMM to model speaker transitions and adopt variational Bayes (VB) inference to estimate the model parameters. From the first row and the second row in Table \ref{tab:unbounded_speakernum}, we can see VBx method is superior to the modularized system in reducing duration errors (DER) and segment-level errors (SER), which also reduces {\em BER}.

EEND takes speaker diarization as a multi-label classification problem and optimizes the speaker diarization label. To alleviate the long-recording issue and arbitrary speaker numbers in EEND, EEND-VC\cite{kinoshita2021advances} first splits the recording into several chucks (30 seconds here). Then, for each chunk, EEND methods, based on permutation invariant training (PIT)\cite{yu2017permutation} training, are used to generate overlap-aware segmentation. Finally, vector clustering (VC) methods are adopted to cluster all utterances. For the experiment of EEND-VC reported here, we do not add any prior information, including the oracle speaker number and oracle segmentation. So the DER score is worse than VBx. Although it is unfair to directly compare the results of those two methods, this comparison can indicate a conflict: for EEND-VC, the DER score is worse, but the SER score is better, shown in the second row and the third row in Table \ref{tab:unbounded_speakernum}. This conflict illustrates that: although EEND-VC does not use oracle segmentation, it is more capable of discriminating segments via PIT and `can-not link' constraints, which reduces {\em SER} and {\em BER}. Besides, in {\em BER}'s second part ($\text{Speaker}^{\text{FA}}$), which means balanced errors caused by false alarm speakers, EEND-VC's is much worse than the modularized system and VBx. This phenomenon is also inconsistent with the fact: the insufficient ability of EEND methods under arbitrary speaker numbers. From the above analysis, our proposed metric, {\em SER} and {\em BER}, can evaluate systems from more perspectives.

\section{Conclusion}
\label{ssec:subhead}

This paper proposes two metrics {\em SER} and {\em BER}. {\em SER} is the segment-level error rate, and {\em BER} is the balanced error rate. Using the connected sub-graph and IoU adaptation strategy, {\em SER} is proposed to accurately solve the segment-matching problem under arbitrary segmentation. Based on {\em SER} and motivated by several conventional metrics, {\em BER} evaluate speaker-weighted, duration, and segment errors in a unified way. With the experiment, {\em BER} shows the potential for emerging algorithms like multi-modal or EEND methods. We hope this metric becomes a valuable tool for the speaker diarization community.

\section{ACKNOWLEDGEMENTS}

This work was supported by State Key Laboratory of Media Convergence Production Technology and Systems Project (No.~SKLMCPTS2020003), Shanghai Municipal Science and Technology Major Project (2021SHZDZX0102), and National Natural Science Foundation of China (Grant No.~92048205).

\vfill\pagebreak

\bibliographystyle{IEEEbib}
\bibliography{strings,refs}

\end{document}